\documentclass{article}
\usepackage{spconf,amsmath,graphicx}
\usepackage{url}
\usepackage{multirow}
\usepackage{threeparttable}
\usepackage{hyperref}

\usepackage{booktabs}

\title{Haha-Pod: An Attempt for Laughter-based Non-Verbal Speaker Verification }
\name{Yuke Lin$^{1,2}$,Xiaoyi Qin$^{1,2}$, Ning Jiang $^{3}$, Guoqing Zhao$^{3}$, Ming Li$^{1,2}$\thanks{Corresponding Author: Ming Li.}}
\address{
$^1$School of Computer Science, Wuhan University, Wuhan, China \\
$^2$Suzhou Municipal Key Laboratory of  Multimodal Intelligent Systems,\\ Duke Kunshan University, Kunshan, China \\
$^3$Mashang Consumer Finance Co., Ltd \\
}
\begin{document}
\maketitle
\begin{abstract}
It is widely acknowledged that discriminative representation for speaker verification can be extracted from verbal speech. However, how much speaker information that non-verbal vocalization carries is still a puzzle. This paper explores speaker verification based on the most ubiquitous form of non-verbal voice, laughter. First, we use a semi-automatic pipeline to collect a new Haha-Pod dataset from open-source podcast media. The dataset contains over 240 speakers' laughter clips with corresponding high-quality verbal speech. Second, we propose a Two-Stage Teacher-Student (2S-TS) framework to minimize the within-speaker embedding distance between verbal and non-verbal (laughter) signals. Considering Haha-Pod as a test set, two trial sets (S2L-Eval) are designed to verify the speaker’s identity through laugh sounds. Experimental results demonstrate that our method can significantly improve the performance of the S2L-Eval test set with only a minor degradation on the VoxCeleb1 test set. The resources for the Haha-Pod dataset can be found at  \url{https://github.com/nevermoreLin/HahaPod}.
\end{abstract}
\section{Introduction}
\label{sec:intro}

In recent years, remarkable advancements have been made in Automatic Speaker Verification (ASV) by extracting deep speaker embeddings with large-scale neural networks such as ECAPA-TDNN~\cite{desplanques20_interspeech} and ResNet~\cite{he2016deep}. Margin-based loss functions like AM-SoftMax~\cite{8331118} and AAM-SoftMax~\cite{deng2019arcface} further enforce higher similarity for intra-class samples and larger distance for inter-class samples. Typically, large-scale datasets comprising numerous speakers' verbal utterances serve as the foundation for speaker verification models and techniques. However, verbal vocal clues for authentication may not be sufficient in certain situations, like tracing the suspect from only non-verbal sounds. For this reason, properly modeling ``garbage” non-verbal vocalizations is a challenging yet interesting task for ASV research (e.g., Catch the \textit{joker} in Gotham through his characteristic laughter). 

Among a couple of non-verbal sounds, the laugh sound occupies a leading proportion under most conversational scenarios~\cite{trouvain2012comparing}. Besides, laughter is commonly associated with a vocal expression of joy——often spelled ``haha" stereotypically. Nevertheless, researchers have noted that the acoustic characteristics of laughter vary significantly across different speakers ~\cite{trouvain2014laughing}, which provides a strong impetus for our research. Consequently, laughter is an appropriate starting point for exploring Non-Verbal Speaker Verification.

Despite the popularity of studies on non-verbal vocalization in other speech processing fields like emotion recognition~\cite{hsu2021speech,huang2019speech} and speech detection~\cite{condron21_interspeech,lea2022nonverbal}, there is a lack of works on the task of non-verbal speaker verification~\cite{janicki2012impact,nandwana15_interspeech,dumpala2017algorithm,dumpala2017improved}. In recent years, Zhang et.al~\cite{zhang2017speaker} employ d-vector to capture embeddings from short-duration trivial events (including laughter). Dumpala et.al~\cite{dumpala2018analysis} incorporate i-vector to perform speaker verification tasks on speech-laughter and laugh utterances. They aim to correctly determine whether two non-verbal utterances belong to the same speaker, yet their methods are conventional ones. However, a more typical scenario is that only the verbal speech is provided as an enrollment utterance, and we need to authenticate the identity by testing on non-verbal segments.

Although some non-verbal vocalization corpora contain various types of non-verbal audio (including laughter)~\cite{janin2003icsi,gemmeke2017audio,gong_vocalsound}, the number of certain non-verbal segments is often limited~\cite{janin2003icsi}, or they do not contain verbal utterances of speakers~\cite{gemmeke2017audio,gong_vocalsound} as enrollment templates. To address this, we build a relatively large-scale dataset comprising both non-verbal and verbal speech from the same speaker. Using laughter as a starting point, we collect a new dataset, named as Haha-Pod, with both laugh segments and the corresponding utterances through our semi-automatic data collection pipeline. Next, two kinds of Speech-Laugh evaluation trial sets (S2L-Eval) are constructed based on Haha-Pod. We have develop a Two-Stage Teacher-Student (2S-TS) technique to further improve the performance in the laughter-based speaker verification scenario with little degradation in the original verbal speech-based situation. The main contribution can be summarized as follows:

\begin{itemize}
\item We introduce a new dataset Haha-Pod which involves both non-verbal laughter and verbal speech segments from 242 speakers.
\item We design two types of evaluation trial sets as benchmarks in the laughter-based speaker verification scenario.
\item We propose a 2S-TS approach to enhance the performance of the S2L-Eval trial sets with minimal impact on the conventional VoxCeleb1 test set.
\end{itemize}

The remaining paper is organized as follows. Section.~\ref{sec:section2} describes the construction pipeline of Haha-Pod and the designing details of the evaluation trial sets. Section.~\ref{sec:method} introduces our proposed 2S-TS Framework. Section.~\ref{sec::result} shows the experimental setup and results. Conclusions are drawn in Section.~\ref{section::Conclusion}.

\section{The Haha-Pod Dataset}
\label{sec:section2}

Given that laughter is a common aspect of human conversation (e.g.phone calls, meetings, and interviews), our objective is to extract laughter fragments from these spontaneous speech signals. For simplicity, we introduce the interview-style podcast in open-source media as our data source.

\subsection{Data Collection Pipeline}
\label{sec:pipeline}

We use a semi-automatic pipeline to assemble our Haha-Pod, as illustrated in Fig.~\ref{fig::pipeline}. The key stages are discussed in detail in the following paragraphs:

\begin{figure}[ht]
  \setlength{\abovecaptionskip}{0.cm}
  \includegraphics[width=0.4\textwidth]{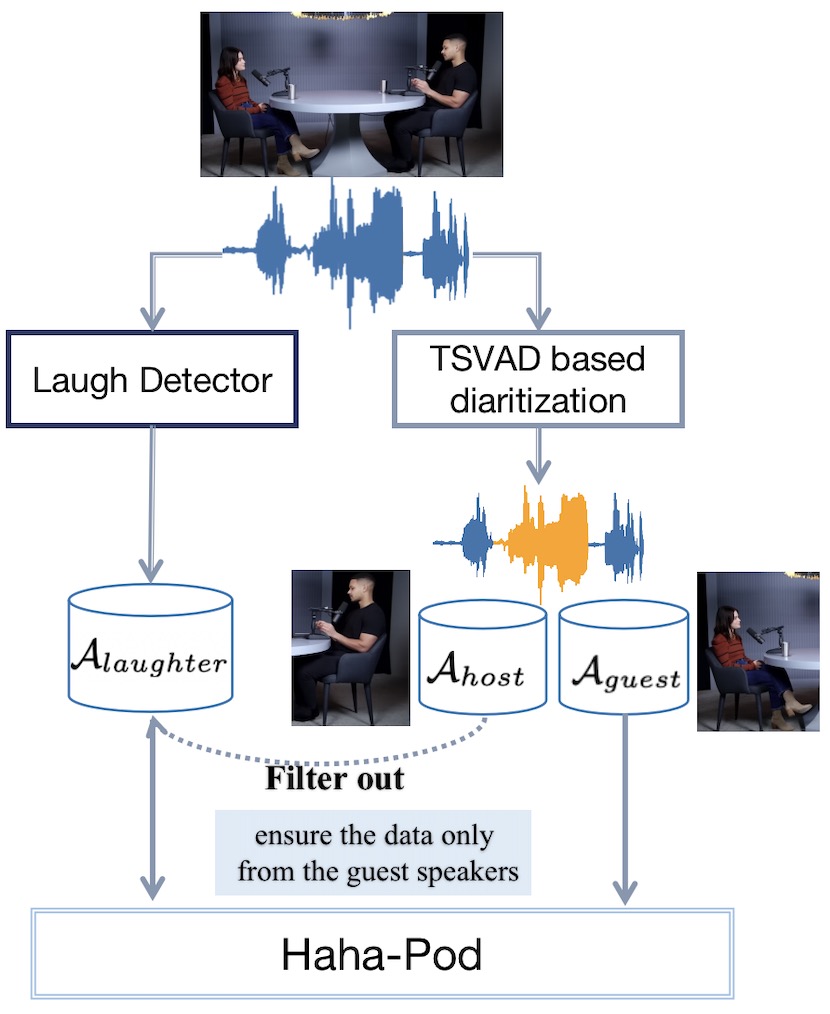}
  \centering
  \caption{{An overview of our data collection pipeline.}}
  \label{fig::pipeline}
  \vspace{-1em}
\end{figure}

\textbf{Stage 1. Candidate Listing.} The initial step is to compile a candidate list of podcast hosts of interest on YouTube. The hosts manage their podcasts and schedule the host-guest interview in many videos they publish. In other words, each host's podcast has a fixed interviewer, while the guests vary with each episode.

\textbf{Stage 2. Audio Downloading.} We proceed to download the audio of these interviews and group them according to the hosts (interviewers). During this phase, we have amassed over 1,000 interview audio from 8 distinct hosts, with a combined duration exceeding 1,200 hours.

\textbf{Stage 3. Laughter Extraction.} We utilize the toolkit proposed by Gillick et al.~\cite{gillick2021robust} to capture laugh segments. Specifically, this resnet-based laughter detector demonstrates excellent robustness in noisy environments. The minimum clip length is set as 1.5 seconds, and the threshold of each frame is 0.5. Subsequently, all the segments that qualify as laughter are defined as $\mathcal{A}_{laughter}$.

\textbf{Stage 4. Host Voice Detection.} Since segments from both the guest and host appear in the interviewing audio, it is imperative to distinguish ``who spoke when.". For this reason, the Sequence-to-Sequence Target-Speaker Voice Activity Detection (Seq2Seq-TSVAD)~\cite{cheng2023target} based diaritization system is implemented to record the timestamps of segments uttered from the hosts and guests. Ten short segments are manually labeled for each host to obtain the target speaker embeddings, which are then fed into the TSVAD model during the inference stage. The accurate time annotations generated by the TSVAD model aid us in retrieving the hosts' segments and subsequently obtaining the guests' speech components from the remaining pieces.

\textbf{Stage 5. Data Filtering.} The speech signals can be segregated into two distinct sets according to the TSVAD system results. Those originating from the hosts are allocated to $\mathcal{A}_{host}$ while those from guests are assigned to $\mathcal{A}_{guest}$. Since we only want the laugh clips from the guests, once a detected laugh clip is accidentally marked as from $\mathcal{A}_{host}$, it is promptly removed from $\mathcal{A}_{laughter}$ to ensures that every clip in $\mathcal{A}_{laughter}$ exclusively pertains to the guest. For the verbal speech signals, we randomly select 15 segments per speaker from $\mathcal{A}_{guest}$. To maintain quality and stability, we only reserve the designated segments for a duration ranging from 5 to 20 seconds.

\textbf{Stage 6. Audio Cropping.} According to the data filtered from $\mathcal{A}_{laughter}$ and $\mathcal{A}_{guest}$ at the previous stage, the segments are cropped from their original interviewing audio and compose our Haha-Pod dataset jointly. In a preliminary sampling, we estimated the laughter accuracy of the Haha-pod to be roughly above 90\% after we set an appropriate threshold.

\subsection{A Glimpse of Data}

\begin{figure}
  \includegraphics[width=0.5\textwidth]{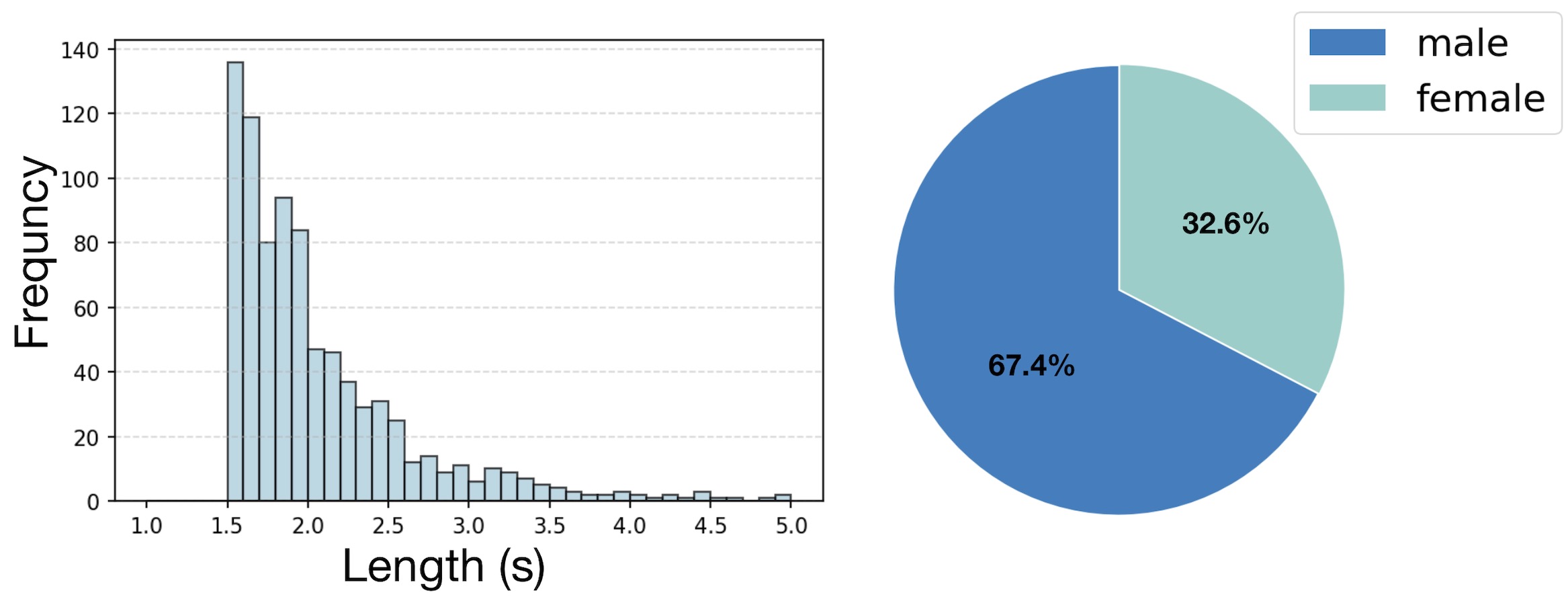}
  \centering
  \caption{{\textbf{Left:} the histogram of duration for laugh segments; \textbf{Right:} the distribution of speaker gender information.}}
  \label{fig::dist}
\end{figure}

Table.~\ref{tab::statistics} displays general statistics while Fig.~\ref{fig::dist} depicts an overview of utterance length and gender distribution. The Haha-Pod dataset consists of 844 laugh segments and 3,630 verbal utterances, contributed by 242 English-speaking individuals across various professions, accents, and age groups. Although our laughter detector has the capability of capturing potential laughter, it is inevitable to have a few speech-with-laugh components. This situation is ignored as it has minimal impact, and further separating these components would be challenging. Notably, the majority of speakers in Haha-Pod are males, accounting for about 67\%. Due to the quiet environment and utilization of high-fidelity microphones during the podcast interviews, the audio quality of the Haha-Pod segments is relatively high. All audio segments are sampled at a frequency of 16 kHz and have undergone the voice activity detection (VAD) module in Librosa~\cite{mcfee2015librosa}.

\begin{table}[ht]\centering \footnotesize
\vspace{-1em}

\caption{\label{tab::statistics} {\it Dataset statistics for Haha-Pod.}}
\begin{tabular}{@{}lcc@{}}
\toprule
\textbf{Types of Utterances} & \textbf{Laughter} & \textbf{ Speech} \\ \midrule
Utterances  & 844 & 3,630 \\
Average duration(s) & 2.092 & 17.361 \\
Average utterances per speaker & 3.48 & 15 \\ \bottomrule
\end{tabular}
\end{table}
\vspace{-1.5em}

\subsection{Trials Construction}
\label{Sec::trial}
The Haha-Pod dataset is well-suited as a test dataset for laughter-based speaker verification. For each testing laugh segment, we randomly select five positive and five negative verbal speech segments as the positive and negative enrollment utterances for trial construction, which we refer to as S2L-Eval-O. Taking into account the high similarity of laugh sounds within each gender, we develop another evaluation protocol named S2L-Eval-H, in which the negative pairs are spoken by different speakers of the same gender. Both evaluation trial sets consist of an equal number of 8,440 pairs.

\section{Methodology}
\label{sec:method}
We propose a novel framework called the Two-Steps Teacher-Student (2S-TS) method, which comprises two distinct procedures.

Our primary objective is to enhance the capability to identify the speaker associated with laughter accurately. However, obtaining a substantial amount of dedicated training data exclusively for laughter and its corresponding speech is challenging, as such data is limited and difficult to simulate. We suggest an alternative approach to address this issue by utilizing the logits output from a laugh detection model. By integrating the laugh detection model into the speaker verification model, we can effectively capture the distinct speaker characteristics of laughter.

Furthermore, to improve the generalization and robustness of the model, we introduce the Teacher-Student (T-S) framework to avoid over-fitting. This method could also bring a closer distance between laughter and speech from the same speaker. The overview of our model training method is illustrated in Fig.~\ref{fig::2S-TSframework}.

\subsection{Stage 1. Laughter-Like Segments Detection}

Due to the inability to obtain laughter for training, we incorporate the ``soft'' result provided by the laugh detector. To capture laughter-like clips at the segment level, we customize a sliding window of 2 seconds, which closely aligns with the average duration of laughter in the Haha-Pod dataset. Since the frozen laugh detector outputs frame-level logits, the sliding window moves along the frames from the beginning and simultaneously calculates the average probability inside. Afterward, we discover a segment with the highest inner-window average probability from each utterance in the training set (e.g., VoxCeleb2). The majority of these selected segments are not genuine laughter, but they do share some similarities with it.

\subsection{Stage 2. Teacher-Student Framework}
Laughter-based speaker verification can be seen as a transfer learning task aiming to achieve high performance not only in laughter-based speaker verification, but also in general speaker verification scenarios. A conventional approach involves fine-tuning the model directly with target domain data (Segments from Stage 1). However, such approach often leads to over-fitting \cite{qin19b_interspeech}. To address this, we propose a simple yet effective method, the Teacher-Student model. This approach leverages the knowledge learned by a generic model to improve the performance of a specific model, ensuring generalization and mitigating over-fitting.

\begin{figure*}
  \setlength{\abovecaptionskip}{0.1cm}
  \includegraphics[width=0.9\textwidth]{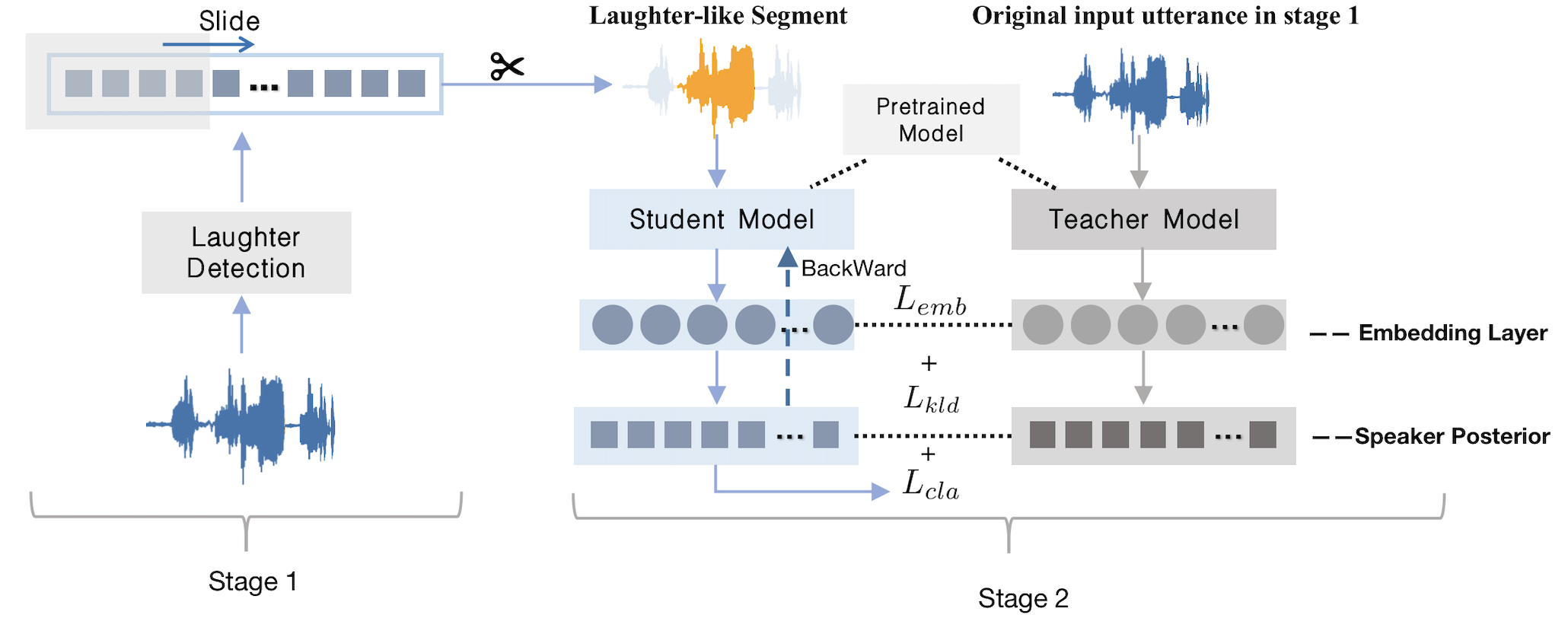}
  \centering
  \caption{{\it An overview of our 2S-TS framework }}
  \label{fig::2S-TSframework}
\end{figure*}

The T-S model has been widely used for robust modeling short-duration utterances~\cite{sang20_interspeech} and far-field utterances~\cite{zhang2021multi} in ASV. As shown in Fig.~\ref{fig::2S-TSframework}, the laughter-like segment with a short duration flows into the student model, and the vanilla utterance is fed into the teacher model concurrently. During the training process, the student model can gradually mimic the probability distribution of the teacher model. Inspired by \cite{sang20_interspeech}, we introduce three objective functions to bridge the gap between one's laughter and verbal speech.

\textbf{Distribution-level Loss.}
Conventionally in T-S learning, the student model learns from the teacher model by minimizing the Kullback-Leibler divergence between the teacher and student output distributions when given parallel data. The loss function is shown as follows:
\vspace{-1em}

{
\begin{equation}
\begin{aligned}
L_{kld} = -\sum_{j=1}^{J}\sum_{i=1}^{I} {P_{t}}(y_i|x_{t,j}) \log({P_{s}}(y_i|x_{s,j}))
\label{con:kld}
\end{aligned}
\end{equation}
}

$i$ and $j$ refer to the speaker and utterance indices, respectively; $x_{t,j}$ and $x_{s,j}$ represent the input to the teacher model and the student model. Since $y_i$ symbolizes $i$-th speaker, $P_t(y_i|\cdot)$ and $P_s(y_i|\cdot)$ provide the speaker $i$-posterior from the head of the teacher model and student model, respectively. With $L_{kld}$, the student model is imposed to generate similar posteriors as the teacher model.

\textbf{Embedding-level Loss.} A discriminate embedding is a powerful representation for identity authentication. Thus, enforcing embedding distance can further improve the laughter-based speaker verification system. Researchers have shown that constraining the similarity of embeddings between paralleled utterances is efficient in T-S learning~\cite{jung2019short,wang2019knowledge}. In our study, We apply cosine-embedding loss as follows: 
{
\begin{equation}
\begin{aligned}
L_{emb} = \frac{1}{N}\sum_{i=1}^{N} (1-\frac{{\bf{z}}^i_s \cdot {\bf{z}}^i_t}{\Vert{\bf{z}}^i_s\Vert \cdot\Vert{\bf{z}}^i_t\Vert})
\label{con:emb}
\end{aligned}
\end{equation}
}
Here, $N$ indicates batch size, and ${\bf{z}}^i_s$ and ${\bf{z}}^i_t$ refer to the $i$-th speaker embedding flowed from the student and teacher model, respectively. By introducing a cosine-distance-based metric, the gap between one's laughter and verbal speech can be further reduced at the embedding level.

\textbf{Speaker-level Loss.} To retain the ability to recognize different speakers, we maintain the ArcFace loss as the speaker-level loss, which also contributes to preventing over-fitting. Eventually, three types of loss are jointly trained with different weights by introducing a multi-task objective function as Eq.~\ref{con::Total}. This joint objective function can help promote both the speaker's discriminative power and the similarity of embeddings between laughter and verbal speech. 
{
\begin{equation}
\begin{aligned}
L_{Total} = \lambda_{cla} \cdot L_{cla} +\lambda_{emb}\cdot L_{emb} +\lambda_{kld}\cdot L_{kld}
\label{con::Total}
\end{aligned}
\end{equation}
}

\section{Experimental Results}
\label{sec::result}
\subsection{Experimental Settings}
\vspace{-0.5em}
\textbf{Dataset.} We train our 2S-TS model on the VoxCeleb2 ~\cite{chung18b_interspeech} dataset, which consists of 2,442 hours of audio from 5,994 speakers. For evaluation, we test on the Haha-Pod dataset with the S2L-Eval trial set and the VoxCeleb1~\cite{nagrani17_interspeech} test set with common trials.

\textbf{Data Processing}. The acoustic features are 80-dimensional log Mel-filterbank energies with a frame length of 25ms and a hop size of 10ms. In our 2S-TS framework, the duration of laughter-like segments and original utterances are truncated to 2s and 5s, respectively.

\textbf{Network.} 
We utilize ResNet34~\cite{cai18_odyssey} as the model backbone, followed by a statistic pooling layer. The widths (channel number) of the residual blocks are \{ 64, 128, 256, 512 \} and a fully connected layer with 256 dimensions followed after the pooling layer is adopted as the speaker embedding layer. The ArcFace (m=0.2, s=32) classifier is introduced to identify speakers.

\textbf{Training Details.} In the pre-training stage, we adopt the on-the-fly data augmentation ~\cite{cai2020fly} and follow a similar training setting in ~\cite{qin2022dku}. During the 1st stage, as shown in Fig.~\ref{fig::2S-TSframework}, the hop length of frames is set to 10ms to calculate laughter-like probability. While in the 2nd stage, The SGD optimizer is employed to update the model parameters and the multi-step learning rate (LR) scheduler with  $10^{-3}$ initial LR drop to  $10^{-6}$ for convergence. The original input utterance will be truncated to 5 seconds to calculate acoustic features. The hyperparameters $\lambda_{cla}$,$\lambda_{emb}$ and $\lambda_{kld}$  are set at 1, 2, 2, respectively.

\textbf{Evaluation Measures.} Cosine similarity is used for trial scoring. Verification performances are measured by Equal Error Rate (EER) and the minimum normalized detection cost function (mDCF) with $P_{target}$ = 0.05.
\vspace{-0.3em}
\subsection{Results and Analysis}
\vspace{-0.5em}

\subsubsection{From Loss Function}

As shown in Table.~\ref{tab::Result_Loss}, the model pre-trained on VoxCeleb2 is used as the baseline for our proposed S2L-Eval trails on Haha-Pod, which obtains degraded performance in the laugh-speech scenario. M2 denotes we use only laughter-like segments from stage 1 as the training data to fine-tune our pre-trained model, while M3 and M4 show ablation results based on different loss compositions. M6 presents the performance of our proposed complete 2S-TS framework, which involves three loss functions from different aspects.

According to the results, optimized with M3 and M4 significantly outperform our baseline M1 in terms of EER at relevant 21.3\% and 21.1\%, respectively. It implies that the consistency constraint on the speaker embeddings and probability distributions between laughter and verbal speech segments of the same speaker is practical. M2 offers an alternative approach that helps the network recognize different types of laughter. Despite achieving 19.7\% and 16.6\% relative EER reductions, M2 fails to enforce the distance between one's laughter and verbal speech. Finally, we yield the best performance on S2L-Eval-O and S2L-Eval-H with the highest EER reduction of 26.9\% and 24.5\% in M6 by composing three types of loss with our 2S-TS framework.
\vspace{-1em}
\begin{table}[ht]\centering \footnotesize
\caption{\label{tab::Result_Loss} {\it Experimental results of S2L-Eval trial sets on Haha-Pod, from the perspective of loss function compositions. }}
\vspace{+0.5em}

\begin{threeparttable}
\begin{tabular}{@{}lccccc@{}}
\toprule
\multirow{2}*{\textbf{ID}} & \multirow{2}*{\textbf{Strategy}} & \multicolumn{2}{c}{\textbf{S2L-Eval-O}} & \multicolumn{2}{c}{\textbf{S2L-Eval-H}}  \\
\cmidrule(lr){3-6} & ~ &  \textbf{EER[\%]} & \textbf{mDCF} & \textbf{EER[\%]} & \textbf{mDCF} \\
\midrule

\textbf{M1} & Pre-trained baseline  & 25.83 & 0.80 & 26.35 & 0.88 \\

\midrule
\textbf{M2} & {$L_{cla}$ (Finetune)}* & 20.88 & 0.69 & 21.97 & 0.74 \\
\textbf{M3} & $L_{cla}$+$L_{emb}$ &  20.31 & 0.67 & 21.43 & 0.72\\
\textbf{M4} & $L_{cla}$+$L_{kld}$ & 20.38 & 0.68 & 21.46 & 0.73 \\
\midrule
\multirow{2}*{\textbf{M6}} & 2S-TS & \multirow{2}*{\textbf{18.89}} & \multirow{2}*{\textbf{0.62}} & \multirow{2}*{\textbf{19.88}} 
 & \multirow{2}*{\textbf{0.66}}\\ & ($L_{cla}$+$L_{emb}$+$L_{kld}$)  \\ 
 \bottomrule

\end{tabular}
\begin{tablenotes}
  \item {* M2 only uses classification loss, which is equivalent to fine-tuning the pre-trained model using target domain data}
\end{tablenotes}
\end{threeparttable}
\end{table}
\vspace{-1em}

\subsubsection{From Student Input}
\vspace{-1em}

\begin{table}[ht]\centering \footnotesize
\caption{\label{tab::Result_Input} {\it Experimental results of S2L-Eval trials on Haha-Pod, from the perspective of different student input. }}
\vspace{+0.5em}

\begin{tabular}{@{}lccccc@{}}
\toprule
\multirow{2}*{\textbf{ID}} & \multirow{2}*{\textbf{Strategy}} & \multicolumn{2}{c}{\textbf{S2L-Eval-O}} & \multicolumn{2}{c}{\textbf{S2L-Eval-H}} \\
\cmidrule(lr){3-6} & ~ &  \textbf{EER[\%]} & \textbf{mDCF} & \textbf{EER[\%]} & \textbf{mDCF} \\
\midrule

\textbf{M1} & Pre-trained baseline  & 25.83 & 0.80 & 26.35 & 0.88 \\
\midrule
\textbf{M5} & Random 2s segment & 20.92 & 0.70 & 22.16 & 0.75
{\vspace{+0.3em}}\\

\multirow{2}*{\textbf{M6}} & 2S-TS & \multirow{2}*{\textbf{18.89}} & \multirow{2}*{\textbf{0.62}} & \multirow{2}*{\textbf{19.88}} 
 & \multirow{2}*{\textbf{0.66}}\\ & (Laughter-like segment)  \\ 

\bottomrule
\end{tabular}
\end{table}

Due to a mismatch in the duration between registered audio and test audio in the testing set of the haha-pod, it makes sense that the duration gap can be bridged by feeding the T-S model with vanilla utterances of varying lengths. As shown in Table.\ref{tab::Result_Input}, we randomly truncate a 2s segment rather than laughter-like segment into the student model.  Even though M2 gains a relative 19\% and 16\% relative EER improvement on S2L-Eval-O and S2L-Eval-H, we can observe that it still can not match M6 because it merely compensates for short utterances rather than laugh segments. 

\subsubsection{From Robustness}

As a problem of Transfer learning, robustness determines its practicability. In our expectation, our proposed system should not only work well on laughter-speech test protocols but also maintain robustness on vanilla speech testing segments. Thus we evaluate our proposed system on the VoxCeleb1 test set. As shown in Table.~\ref{tab::VoxResult}, in terms of EER, we can discover that directly fine-tuning the pre-trained model with laughter-like segments results in a severe degradation on the Vox1 test set. In detail, the performances of Vox-O, Vox-E, and Vox-H decrease by 6.5\%, 8.5\% and 19.5\%, respectively. Our 2S-TS framework makes little side-effects (\textless 3\% relatively) on both Vox-O and Vox-E and a tolerable degradation (\textless 6\% relatively) on Vox-H. This result interprets that the T-S framework can effectively alleviate over-fitting. Furthermore, the result suggests the robustness of our 2S-TS framework and proves its application potential.

\vspace{-0.8em}

\begin{table}[ht]
\centering
\caption{\label{tab::VoxResult} {\it Experimental results on VoxCeleb1 test set.}}
\vspace{+0.5em}
\begin{tabular}{@{}ccccc@{}}
\toprule
{\textbf{ID}} & \multicolumn{1}{c}{\textbf{Strategy}} & \textbf{Vox-O} & \textbf{ Vox-E} & \textbf{ Vox-H} \\ \midrule
\textbf{M1} & Pre-trained baseline & 0.86 & 1.14 & 1.94 \vspace{+0.3em}\\

\textbf{M2} & $L_{cla}$ (Finetune)  & 0.91 & 1.23 & 2.32
\vspace{+0.4em}\\
\textbf{M6} & 2S-TS & 0.87 & 1.16 & 2.06 \\
\bottomrule
\end{tabular}
\end{table}

\vspace{-2em}

\subsection{Visualization}

\begin{figure}[ht]
  \setlength{\abovecaptionskip}{0.1cm}
  \includegraphics[width=0.48\textwidth]{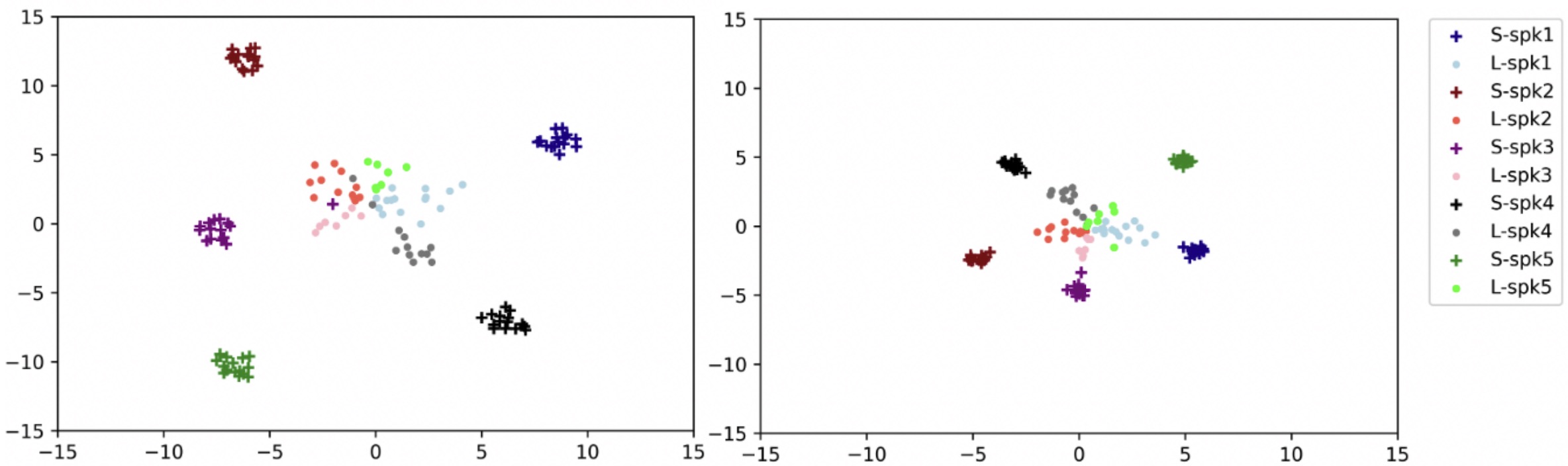}
  \centering
  \caption{{A figure of 2D embedding (after T-SNE) comparison between M1 (Left) and M6 (Right). ``L-" and ``S-" in the legend correspond to the embeddings derived from the non-verbal laugh segments or the verbal speech, respectively.}}
  \label{fig::tsne}
\end{figure}

To further demonstrate the effectiveness of our method, we extract all laugh segments and verbal speech from a randomly selected group of five speakers in Haha-Pod. After extracting the speaker embedding from two distinct models, we adopt t-SNE to visualize the distribution of embeddings in 2-D space in Fig.~\ref{fig::tsne}.

\label{sec::figure}
\vspace{-1em}

When using a pre-trained model (M1) as the feature extractor, it can be discovered that there is a relatively large distance between the embeddings of one's laughter and verbal utterance. Upon applying our 2S-TS framework (M6), we can observe that the intra-class distance is enforced, which indicates a higher similarity of speaker embedding between non-verbal laughter and verbal speech from the same speaker.

Furthermore, a clustering effect can be found among all the laughter embeddings, suggesting that different speakers' laughter may share certain commonalities, such as contents and pitches. Nonetheless, due to the limited personalized information conveyed by laughter, a distinct performance gap still exists between non-verbal laughter and verbal speech. This indirectly proves the challenge of using laughter to verify human voices, which needs further investigation

\vspace{-0.5em}
\section{Conclusion}
\label{section::Conclusion}

Non-verbal speaker verification is an emerging topic of interest with potential applications. In this paper, we present our approach to laughter-based non-verbal speaker verification. We begin by describing a semi-automatic data collection pipeline that allows us to construct a new dataset called Haha-Pod, which contains both verbal speech and non-verbal laughter. Next, we introduce a two-stage teacher-student framework that enables us to minimize the distance between the embeddings of non-verbal laughter and verbal speech from the same speaker. Experimental results and ablation study demonstrate that our method can substantially boost laughter-based non-verbal speaker verification performance without sacrificing performance in the normal speech-to-speech testing scenario.
\vspace{-0.5em}

\section{Acknowledgement}
\label{section::acknowledgement}
This research is funded in part by the National Natural Science Foundation of China (62171207), Science and Technology Program of Suzhou City(SYC2022051) and MaShang Consumer Finance Co.Ltd. Many thanks for the computational resource provided by the Advanced Computing East China Sub-Center.

\newpage
\bibliographystyle{IEEEbib}
\bibliography{refs}

\begin{thebibliography}{10}

\bibitem{desplanques20_interspeech}
Brecht Desplanques, Jenthe Thienpondt, and Kris Demuynck,
\newblock ``{ECAPA-TDNN: Emphasized Channel Attention, Propagation and
  Aggregation in TDNN Based Speaker Verification},''
\newblock in {\em Proc. Interspeech}, 2020, pp. 3830--3834.

\bibitem{he2016deep}
Kaiming He, Xiangyu Zhang, Shaoqing Ren, and Jian Sun,
\newblock ``Deep residual learning for image recognition,''
\newblock in {\em Proc. CVPR}, 2016, pp. 770--778.

\bibitem{8331118}
Feng Wang, Jian Cheng, Weiyang Liu, and Haijun Liu,
\newblock ``Additive margin softmax for face verification,''
\newblock {\em IEEE Signal Processing Letters}, vol. 25, no. 7, pp. 926--930,
  2018.

\bibitem{deng2019arcface}
Jiankang Deng, Jia Guo, Niannan Xue, and Stefanos Zafeiriou,
\newblock ``Arcface: Additive angular margin loss for deep face recognition,''
\newblock in {\em Proc.CVPR}, 2019, pp. 4690--4699.

\bibitem{trouvain2012comparing}
J{\"u}rgen Trouvain and Khiet~P Truong,
\newblock ``Comparing non-verbal vocalisations in conversational speech
  corpora,''
\newblock in {\em Proc.the LREC Workshop on Corpora for Research on Emotion
  Sentiment and Social Signals}, 2012, pp. 36--39.

\bibitem{trouvain2014laughing}
J{\"u}rgen Trouvain,
\newblock ``Laughing, breathing, clicking-the prosody of nonverbal
  vocalisations,''
\newblock in {\em Proc.Speech Prosody}, 2014, vol.~7, pp. 598--602.

\bibitem{hsu2021speech}
Jia-Hao Hsu, Ming-Hsiang Su, Chung-Hsien Wu, and Yi-Hsuan Chen,
\newblock ``Speech emotion recognition considering nonverbal vocalization in
  affective conversations,''
\newblock {\em IEEE/ACM Transactions on Audio, Speech, and Language
  Processing}, vol. 29, pp. 1675--1686, 2021.

\bibitem{huang2019speech}
Kun-Yi Huang, Chung-Hsien Wu, Qian-Bei Hong, Ming-Hsiang Su, and Yi-Hsuan Chen,
\newblock ``Speech emotion recognition using deep neural network considering
  verbal and nonverbal speech sounds,''
\newblock in {\em Proc.ICASSP}, 2019, pp. 5866--5870.

\bibitem{condron21_interspeech}
Scott Condron, Georgia Clarke, Anita Klementiev, Daniela Morse-Kopp, Jack
  Parry, and Dimitri Palaz,
\newblock ``{Non-Verbal Vocalisation and Laughter Detection Using
  Sequence-to-Sequence Models and Multi-Label Training},''
\newblock in {\em Proc.Interspeech}, 2021, pp. 2506--2510.

\bibitem{lea2022nonverbal}
Colin Lea, Zifang Huang, Dhruv Jain, Lauren Tooley, Zeinab Liaghat, Shrinath
  Thelapurath, Leah Findlater, and Jeffrey~P Bigham,
\newblock ``Nonverbal sound detection for disordered speech,''
\newblock in {\em Proc.ICASSP}, 2022, pp. 7397--7401.

\bibitem{janicki2012impact}
Artur Janicki,
\newblock ``On the impact of non-speech sounds on speaker recognition,''
\newblock in {\em Proc. TSD}, 2012, pp. 566--572.

\bibitem{nandwana15_interspeech}
Mahesh~Kumar Nandwana, Hynek Bořil, and John H.~L. Hansen,
\newblock ``{A new front-end for classification of non-speech sounds: a study
  on human whistle},''
\newblock in {\em Proc.Interspeech}, 2015, pp. 1982--1986.

\bibitem{dumpala2017algorithm}
Sri~Harsha Dumpala and KNRK Alluri,
\newblock ``An algorithm for detection of breath sounds in spontaneous speech
  with application to speaker recognition,''
\newblock in {\em Proc.SPECOM}, 2017, pp. 98--108.

\bibitem{dumpala2017improved}
Sri~Harsha Dumpala and Sunil~Kumar Kopparapu,
\newblock ``Improved speaker recognition system for stressed speech using deep
  neural networks,''
\newblock in {\em Proc.IJCNN}, 2017, pp. 1257--1264.

\bibitem{zhang2017speaker}
Miao Zhang, Yixiang Chen, Lantian Li, and Dong Wang,
\newblock ``Speaker recognition with cough, laugh and" wei",''
\newblock in {\em Proc.APSIPA ASC}. IEEE, 2017, pp. 497--501.

\bibitem{dumpala2018analysis}
Sri~Harsha Dumpala, Ashish Panda, and Sunil~Kumar Kopparapu,
\newblock ``Analysis of the effect of speech-laugh on speaker recognition
  system.,''
\newblock in {\em Proc.Interspeech}, 2018, pp. 1751--1755.

\bibitem{janin2003icsi}
Adam Janin, Don Baron, Jane Edwards, Dan Ellis, David Gelbart, Nelson Morgan,
  Barbara Peskin, Thilo Pfau, Elizabeth Shriberg, Andreas Stolcke, et~al.,
\newblock ``The icsi meeting corpus,''
\newblock in {\em Proc.ICASSP}, 2003, vol.~1, pp. I--I.

\bibitem{gemmeke2017audio}
Jort~F Gemmeke, Daniel~PW Ellis, Dylan Freedman, Aren Jansen, Wade Lawrence,
  R~Channing Moore, Manoj Plakal, and Marvin Ritter,
\newblock ``Audio set: An ontology and human-labeled dataset for audio
  events,''
\newblock in {\em Proc.ICASSP}, 2017, pp. 776--780.

\bibitem{gong_vocalsound}
Yuan Gong, Jin Yu, and James Glass,
\newblock ``Vocalsound: A dataset for improving human vocal sounds
  recognition,''
\newblock in {\em Proc.ICASSP}, 2022, pp. 151--155.

\bibitem{gillick2021robust}
Jon Gillick, Wesley Deng, Kimiko Ryokai, and David Bamman,
\newblock ``Robust laughter detection in noisy environments.,''
\newblock in {\em Proc.Interspeech}, 2021, pp. 2481--2485.

\bibitem{cheng2023target}
Ming Cheng, Weiqing Wang, Yucong Zhang, Xiaoyi Qin, and Ming Li,
\newblock ``Target-speaker voice activity detection via sequence-to-sequence
  prediction,''
\newblock in {\em Proc.ICASSP}. IEEE, 2023, pp. 1--5.

\bibitem{mcfee2015librosa}
Brian McFee, Colin Raffel, Dawen Liang, Daniel~P Ellis, Matt McVicar, Eric
  Battenberg, and Oriol Nieto,
\newblock ``librosa: Audio and music signal analysis in python,''
\newblock in {\em Proc.SciPy}, 2015, vol.~8, pp. 18--25.

\bibitem{qin19b_interspeech}
Xiaoyi Qin, Danwei Cai, and Ming Li,
\newblock ``{Far-Field End-to-End Text-Dependent Speaker Verification Based on
  Mixed Training Data with Transfer Learning and Enrollment Data
  Augmentation},''
\newblock in {\em Proc. Interspeech 2019}, 2019, pp. 4045--4049.

\bibitem{sang20_interspeech}
Mufan Sang, Wei Xia, and John~H.L. Hansen,
\newblock ``{Open-Set Short Utterance Forensic Speaker Verification Using
  Teacher-Student Network with Explicit Inductive Bias},''
\newblock in {\em Proc.Interspeech}, 2020, pp. 2262--2266.

\bibitem{zhang2021multi}
Li~Zhang, Qing Wang, Kong~Aik Lee, Lei Xie, and Haizhou Li,
\newblock ``Multi-level transfer learning from near-field to far-field speaker
  verification,''
\newblock {\em arXiv preprint arXiv:2106.09320}, 2021.

\bibitem{jung2019short}
Jee-weon Jung, Hee-Soo Heo, Hye-jin Shim, and Ha-Jin Yu,
\newblock ``Short utterance compensation in speaker verification via
  cosine-based teacher-student learning of speaker embeddings,''
\newblock in {\em Proc.ASRU workshop}, 2019, pp. 335--341.

\bibitem{wang2019knowledge}
Shuai Wang, Yexin Yang, Tianzhe Wang, Yanmin Qian, and Kai Yu,
\newblock ``Knowledge distillation for small foot-print deep speaker
  embedding,''
\newblock in {\em Proc.ICASSP}, 2019, pp. 6021--6025.

\bibitem{chung18b_interspeech}
Joon~Son Chung, Arsha Nagrani, and Andrew Zisserman,
\newblock ``{VoxCeleb2: Deep Speaker Recognition},''
\newblock in {\em Proc.Interspeech}, 2018, pp. 1086--1090.

\bibitem{nagrani17_interspeech}
Arsha Nagrani, Joon~Son Chung, and Andrew Zisserman,
\newblock ``{VoxCeleb: A Large-Scale Speaker Identification Dataset},''
\newblock in {\em Proc.Interspeech}, 2017, pp. 2616--2620.

\bibitem{cai18_odyssey}
Weicheng Cai, Jinkun Chen, and Ming Li,
\newblock ``{Exploring the Encoding Layer and Loss Function in End-to-End
  Speaker and Language Recognition System},''
\newblock in {\em Proc.Odyssey}, 2018, pp. 74--81.

\bibitem{cai2020fly}
Weicheng Cai, Jinkun Chen, Jun Zhang, and Ming Li,
\newblock ``On-the-fly data loader and utterance-level aggregation for speaker
  and language recognition,''
\newblock {\em IEEE/ACM Transactions on Audio, Speech, and Language
  Processing}, vol. 28, pp. 1038--1051, 2020.

\bibitem{qin2022dku}
Xiaoyi Qin, Na~Li, Yuke Lin, Yiwei Ding, Chao Weng, Dan Su, and Ming Li,
\newblock ``The dku-tencent system for the voxceleb speaker recognition
  challenge 2022,''
\newblock {\em arXiv preprint arXiv:2210.05092}, 2022.

\end{thebibliography}

\end{document}